\documentclass[prb,twocolumn,showpacs]{revtex4-1}

\usepackage{graphicx}
\usepackage{multirow}
\usepackage{dcolumn}
\usepackage{bm}
\usepackage[normalem]{ulem}
\usepackage{amsmath}
\usepackage{amsfonts}
\usepackage{amssymb}
\usepackage{color}
\usepackage{colordvi}
\usepackage{dcolumn}

\begin{document}

\title{Effect of magnetic field on competition between
superconductivity and charge order below the pseudogap state}
\date{\today }
\author{ H. Meier$^{1,2}$, M. Einenkel$^{1,2}$, C. P\'epin$^{2}$, and K. B.
Efetov$^{1,2}$ }
\affiliation{$^1$Institut f\"ur Theoretische Physik III, Ruhr-Universit\"at Bochum, 44780
Bochum, Germany\\
$^2$IPhT, CEA-Saclay, L'Orme des Merisiers, 91191 Gif-sur-Yvette, France }

\begin{abstract}
We theoretically investigate the $T$-$B$ phase diagram of cuprates building
on the $\mathrm{SU}(2)$ pseudogap order recently obtained from the
spin-fermion model. At low temperatures, our analysis reveals an almost
temperature-independent critical magnetic field~$B_{\mathrm{co}}$ at which $d$-wave
superconductivity switches to a phase of charge order. For temperatures beyond a certain value of the order of the
zero-field superconducting transition temperature~$T_c$, the critical
field~$B_{\mathrm{co}}(T)$ sharply grows. We compare our results derived
within the effective $\mathrm{SU}(2)$ non-linear $\sigma$-model with the
phase diagram recently obtained by sound velocity measurements by LeBoeuf
\emph{et al.} [Nat.~Phys.~\textbf{9}, 79~(2013)].
\end{abstract}

\maketitle

%%%%%%%%%%%%%%%%%%%%%%%%%%%%%%%%%%%%%%%
%
% title, authors, abstract
%
%%%%%%%%%%%%%%%%%%%%%%%%%%%%%%%%%%%%%%%

%%%%%%%%%%%%%%%%%%%%%%%%%%%%%%%%%%%%%%%
%
% introduction
%
%%%%%%%%%%%%%%%%%%%%%%%%%%%%%%%%%%%%%%%

\section{Overview}
After more than 25 years of investigation, high-$T_{c}$ cuprate
superconductors still retain a lot of their mystery. \cite{leenagaosawen} At the same
time, experiments quarried a huge body of unprecedented data to fuel the
theoretical investigation. Among all the phases these compounds exhibit, the
normal phase and especially the infamous \textquotedblleft
pseudogap\textquotedblright\ phase remain particularly controversial.
The proximity to antiferromagnetism, the vicinity of a metal-insulator Mott
transition, pre-formed superconducting pairs, or the presence of hidden
competing orders are exemplary scenarios that have been suggested to explain
the curious depression in most thermodynamic and transport quantities within
the pseudogap. \cite{Norman:605918}

Recently, a series of ground-breaking
experiments has revived the debate about the presence of a modulated state in competition
with superconductivity. While in La-based superconductors, the presence of static
modulated spin-charge orders at commensurate doping close to $1/8$ is
now well-established \cite{tranquada}, the situation for YBCO and BSCCO compounds
is much less clear. First, if we assume a stripe order, it is most probably
of dynamical nature in these compounds, and second, it has been suggested
that the modulation in those compounds is biaxial, of the form of a
checkerboard as reported by scanning tunneling microscopies (STM) in Bi$_{2}$Sr$_{2}$CaCu$_{2}$O$_{8+\delta }$. \cite{wise,fujita}
In contrast to the stripe structure, the checkerboard
structure has been claimed to be incommensurate with the Cu-O lattice, with a
periodicity determined by wave vectors close to those connecting the
nearest antinodes.

Soft \cite{ghiringhelli,achkar} and hard X-ray \cite{chang,blackburn}
measurements confirmed these findings, disclosing the presence of an
incommensurable biaxial purely charge structure in YBa$_{2}$Cu$_{3}$O$_{6+x}$
in the doping range $0.09\leq p\leq 0.13$, with modulation wave vectors
compatible with those observed by STM. Quantum oscillation measurements \cite{DoironLeyraud:2007bj,laliberte:2011}
associated with a negative Seebeck coefficient and negative Hall constant
point out the existence of at least one small electron pocket at low
temperatures \cite{Chang:2010,Millis:2007wx,Sebastian:2012vd,Sebastian:2012ip}. The Fermi surface reconstruction implied by these
measurements may be attributed to the proximity of a competing modulated order. This interpretation is in line with the results of earlier
nuclear magnetic resonance (NMR) measurements \cite{wu,wu2013} indicating the existence of a modulated charge order, though
not being decisive yet about whether the form is stripes or checkerboard and commensurate or incommensurate.

The charge order observed in these experiments appears or is enhanced under the application
of an external magnetic field~$B$. \cite{chang,wu,leboeuf} A
recent study \cite{leboeuf} of the $T$-$B$ phase diagram of underdoped YBa$_{2}$Cu$_{3}$O$_{6+x}$ with
$p=0.108$ has revealed interesting features and convincing evidence for a two-dimensional charge
modulation. Applying magnetic fields up to~$30\ \mathrm{T}$, the authors of Ref.~\onlinecite{leboeuf}
extracted from ultra-sound measurements the elastic constants of the
compound, which allowed to identify the phase transitions. Below a critical temperature $T_{\mathrm{co}}\approx 40\ \mathrm{K}$,
a static charge order stabilizes abruptly above a critical field of $B_{\mathrm{co}}\approx 18\ \mathrm{T}$
that is temperature-independent (see Fig.~2 of Ref.~\onlinecite{leboeuf}). Contrarily, for magnetic fields $B>B_{\mathrm{co}}$,
the boundary of the static charge order practically remains of order~$T_{\mathrm{co}}$, only weakly depending on
the magnetic field.

In a recent theoretical work \cite{emp} on the antiferromagnet-normal metal quantum critical point (QCP),
several of us have come to a conclusion about the existence of a pseudogap
phase in its vicinity that is characterized by a composite $\mathrm{SU}(2)$
order parameter, combining the $d$-wave
superconducting and a particle-hole suborder. The latter is a charge order characterized by a spatially modulated electric quadrupolar moment
and may thus be referred to as a quadrupole density wave. The degeneracy
between superconductivity and charge order is lifted for a finite Fermi surface curvature, resulting in the stabilization of superconductivity at low
temperatures $T<T_c$. At higher temperatures~$T_c<T<T^*$ ---the pseudogap region--- thermal fluctuations described by an $\mathrm{SU}(2)$ non-linear $\sigma $-model are strong and prevent the two
suborders from disentangling. Here, they are mixed together but
fluctuations destroy any long-range order.

In this paper, we generalize the non-linear $\sigma$-model derived in Ref.~\onlinecite{emp}
and include an external magnetic field~$B$ affecting the orbital motion. As the
electron pairing is purely singlet, effects of the magnetic field
on the electron spins are negligible. The magnetic field~$B$ favors the charge suborder of the
pseudogap. Sufficiently strong fields~$B>B_0$ may even surmount the
curvature threshold and establish a charge order at low temperatures. The system is thus allowed
to switch between the two suborders, superconductivity and charge order, which is
controlled by the strength of the applied field.
At the critical field~$B_0$, the system is degenerate between
superconducting and particle-hole order states. The fact that such a simple
switching mechanism might be at the heart of the physics of the pseudogap
state in the cuprates is remarkable and may even have the
potential to restrain the window of validity of theoretical interpretations.

\begin{figure}[h]
\centerline{\includegraphics[width=0.9\linewidth]{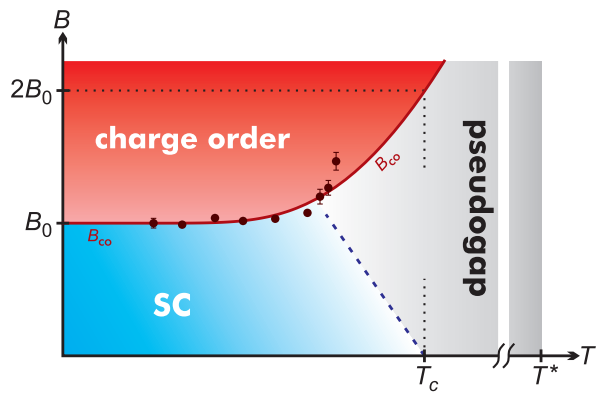}}
\caption{(Color online) $T$-$B$ phase diagram following our analysis of the
fluctuations around the mean-field order parameter of the spin-fermion
model. The dark red dots
are from the sound velocity measurements reported in Ref.~\onlinecite{leboeuf} with~$B_0 \approx 18\ \mathrm{T}$
and~$T_c \approx 60.7\ \mathrm{K}$ while the red curve~$B_{\mathrm{co}}(T)$, Eq.~(\ref{k6}), has been fitted to the experimental data
with the constraint that $B_{\mathrm{co}}(T_c)=2B_0$.
}
\label{fig1}
\end{figure}

Our main results are illustrated by Fig.~\ref{fig1} representing in
the $T$-$B$ plane the regions of charge order, $d$-wave superconductivity (SC), and the
pseudogap state. We see that the border between the charge order and the SC
state is flat at low temperatures while the charge order--and SC--pseudogap borders
depend on the magnetic field~$B$ only logarithmically. This picture agrees well
with the experimental $T$-$B$~phase diagram of Ref.~\onlinecite{leboeuf}.

%%%%%%%%%%%%%%%%%%%%%%%%%%%%%%%%%%%%%%%
%
% phase diagram
%
%%%%%%%%%%%%%%%%%%%%%%%%%%%%%%%%%%%%%%%
%%%%%%%%%%%%%%%%%%%%%%%%%%%%%%%%%%%%%%%

\section{Model for the pseudogap state}
We investigate the pseudogap state in the cuprate superconductors
beginning with an effective spin-fermion model (see, e.g., Refs.~\onlinecite{acs,ms2}) describing
electrons coupled to quantum critical antiferromagnetic paramagnons. The
Lagrangian for the $(2\!+\!1)$-dimensional model is written as
\begin{align}
\mathcal{L} &=
 \chi^{\dagger}\
 \big(\hbar\partial_{\tau } + \varepsilon(-\mathrm{i}\hbar\nabla) + \lambda\vec{\phi}\vec{\sigma}
 \big)\ \chi \ .  \label{2a01}
\end{align}
The field~$\vec{\phi}$ describes the paramagnons that couple to the spin~$\vec{\sigma}$
of the electronic fields~$\chi$. The paramagnon excitations
are modeled by the correlation function
\begin{align}
\langle\phi_{\omega ,\mathbf{k}}^{i}\phi _{-\omega ,-\mathbf{k}}^{j}\rangle
 \propto \frac{\delta _{ij}}{(\omega /v_{s})^{2}+(\mathbf{k}-\mathbf{Q})^{2}+a}
 \label{k1}
\end{align}
where~$v_{s}$ is the wave velocity and~$\mathbf{Q}$ the antiferromagnetic
ordering vector below the QCP. The distance to the QCP is controlled by the
parameter~$a$ with the QCP itself situated at~$a=0$. In this study, we consider the proximity
of the QCP to its right ($a\geq 0$) but, at finite temperatures, the results should also
apply qualitatively to the near quantum critical region on its left ($a<0$). Landau damping modifies
the paramagnon propagator but this effect is reduced when the pseudogap opens. \cite{emp}

The mean-field analysis \cite{emp} of the spin-fermion model~(\ref{2a01})
indicates that below a temperature~$T^*$, orders in both the 
superconducting and particle-hole channel emerge and combine to form a
composite order parameter~$\mathcal{O}(\varepsilon)=b(\varepsilon)u$ with $b(\varepsilon)$
a function of fermionic Matsubara frequencies and $u$ denoting an $\mathrm{SU}(2)$
matrix in the Gor'kov-Nambu particle-hole space. The typical
scale of~$b(\varepsilon)$ is $k_{B}T^*$.
The matrix $u$ can be parametrized by two complex
order parameters~$\Delta_{+}$ and~$\Delta_{-}$ for superconducting and
charge suborders, respectively,
\begin{align}
u=\left(
\begin{array}{cc}
\Delta _{-} & \Delta _{+} \\
-\Delta _{+}^{\ast } & \Delta _{-}^{\ast } \\
\end{array}
\right)\label{3a01}
\end{align}
while unitarity imposes the constraint~$|\Delta _{+}|^{2}+|\Delta _{-}|^{2}=1$.

As shown in Ref.~\onlinecite{emp}, fluctuations around a
particular mean-field solution are accurately described in terms of a
two-dimensional $\mathrm{SU}(2)$ non-linear $\sigma$-model. At temperatures~$T>0$, the partition function for the low-lying Goldstone modes
has the form~$\mathcal{Z}=\int\exp(-\mathcal{F})\mathcal{D}u$ with
\begin{align}
\mathcal{F}=
\frac{1}{t}
\int\mathrm{tr}
\big[
   \nabla u^{\dagger }\nabla u + \kappa ^{2}u^{\dagger }\tau _{3}u\tau _{3}
\big]\
\mathrm{d}^{2}\mathbf{r}\ .
\label{3a02}
\end{align}
In terms of microscopic parameters, $t=(8\pi /J_{1}\sin\delta)\ (k_{B}T/\hbar vS)$,
where $v$ is the Fermi velocity, $S$ the size of
a gapped hot spot on the Fermi surface, and $\delta$ the
angle between the Fermi velocities at two hot spots connected by the
ordering vector~$\mathbf{Q}$. \cite{footnote01} For~$T<T^*$, $J_1 \sim \bar{J_1} \approx 0.25$ whereas approaching $T^*$, $J_1$ turns to zero
(cf. the Supplemental Material of Ref.~\onlinecite{emp}). We may estimate~$T^*$
by $k_B T^* = \alpha^{-1}\ (2\bar{J_1}\sin\delta/3)\ \hbar v S$ with~$\alpha\sim 1$.
The coupling constant~$t = (16\pi/3)\ (T/\alpha T^*)$ is the effective temperature
in the $\sigma$-model.

The coupling constant~$\kappa^{2}$ of the second term in Eq.~(\ref{3a02})
emerges from curvature corrections to the linearized spin-fermion model. On the right of the QCP,
$\kappa^{2}$ is enhanced with growing~$a>0$.
Containing the third Pauli matrix~$\tau_{3}$ in the Gor'kov-Nambu space,
the second term breaks the~$\mathrm{SU}(2)$ rotational symmetry between the particle-hole
and superconducting suborders to favor the latter. The renormalization group (RG) analysis
shows that superconductivity is stablized below a temperature~$T_c<T^*$
determined by the magnitude of~$\kappa$. The transition between
superconductivity and the disordered pseudogap phase is driven by thermal
fluctuations.
The $\mathrm{SU}(2)$ symmetry of the linearized spin-fermion model has first been noticed in Ref.~\onlinecite{ms2}
where the particle-hole order was discussed as a subleading instability in the bond correlation functions.

In order to investigate the $T$-$B$ phase diagram, we insert the magnetic field~$B$ into the non-linear $\sigma $-model~(\ref{3a02})
by making the replacement
\begin{align}
\nabla u \rightarrow
\nabla u + \mathrm{i}e\mathbf{\bar{A}}[\tau _{3},u]
\label{k2}
\end{align}
where $[\cdot ,\cdot ]$ is the commutator. In Landau gauge,
$\mathbf{\bar{A}}(x,y)=-\tfrac{1}{2}\sin\delta\ By\ \mathbf{e}_{x}$ with~$\mathbf{e}_{x}$
the unit vector in the $x$-direction. The unusual prefactor of $\tfrac{1}{2}\sin\delta$
in the ``reduced vector potential''~$\mathbf{\bar{A}}$ is due to a change of coordinates
during the derivation of Eq.~(\ref{3a02}), cf. Ref.~\onlinecite{emp}.

%%%%%%%%%%%%%%%%%%%%%%%%%%%%%%%%%%%%%%%
%
% Bc2
%
%%%%%%%%%%%%%%%%%%%%%%%%%%%%%%%%%%%%%%%

\section{$T$--$B$ phase diagram}

\paragraph*{Charge order--SC transition.}
Let us first consider the transition between superconductivity (SC) and the charge suborder at low temperatures~$T\ll
T^*$. The critical field~$B_{\mathrm{co}}$
of this transition is at the same time the field~$B_{c2}$ of the type-II
superconductor. At very strong magnetic fields, SC clearly may not exist so that the charge order prevails instead. In order to look
for the first appearance of the superconducting order parameter when lowering the magnetic field~$B$,
we may assume that $|\Delta_{+}|$, cf. Eq.~(\ref{3a01}), is small and expand the functional~$\mathcal{F}$ for the free energy
up to quadratic terms in $\Delta_{+}$,
\begin{align}
\mathcal{F}[\Delta _{+},\chi ] & \simeq
\frac{1}{t} \int
   \big\{
      \big|(\nabla +2\mathrm{i}e\mathbf{\bar{A}})\Delta _{+}
      \big|^{2}
      -2 \kappa ^{2} \big|\Delta _{+}\big|^{2}  \notag \\
    & \quad \quad \quad + (\nabla \chi) ^{2} \big(1 - \big|\Delta _{+}\big|^{2}\big)
   \big\}
   \ \mathrm{d}^{2}\mathbf{r} \label{3b02}
\end{align}
with $\chi$ the phase of the charge order parameter~$\Delta_{-}$.

This quadratic form is strictly positive in the charge order state but loses its
stability at $B < B_{\mathrm{co}}$. At low temperatures, we can neglect the
$(\nabla \chi )^{2}$-term. Then the study of the approximated functional
in Eq.~(\ref{3b02}) is reduced to solving the Schr\"{o}dinger
equation in the presence of a magnetic field. As a result, we find that non-zero solutions
for $\Delta _{+}$ appear below the critical magnetic field~$B_{\mathrm{co}}= B_0$ with
\begin{align}
  B_0 = \frac{\hbar \kappa ^{2}}{e\sin \delta }\ .
\label{k3}
\end{align}
Below $B_0$, we expect the formation of superconducting
vortices that in contrast to the usual scenario must contain charge order cores instead of normal metal phases.
The details of the physics in this region of the phase diagram and, in particular, the question of coexistence of charge and superconducting
orders at low temperatures \cite{wu2013,taillefer2013} will be left for a separate study.

Equation~(\ref{k3}) shows that at small~$T$, where thermal fluctuations are weak,
the critical field $B_{\mathrm{co}}$ does not depend on temperature. It also relates the value of
the coupling constant~$\kappa^{2}$ of the $\sigma $-model~(\ref{3a02}) to the
experimentally measurable magnetic field~$B_0$ that separates the two phases.

%%%%%%%%%%%%%%%%%%%%%%%%%%%%%%%%%%%%%%%
%
% RG analysis
%
%%%%%%%%%%%%%%%%%%%%%%%%%%%%%%%%%%%%%%%

\paragraph*{SC--pseudogap and charge order--pseudogap transitions.}
In the absence of a magnetic field, superconductivity is stabilized at
temperatures $T<T_c$ whereas at $T>T_{c}$ thermal fluctuations destroy a long-range order in the pseudogap. \cite{emp} If
due to a strong magnetic field the ground state is a charge order, we may expect a similar transition
between pseudogap and this order, which is stabilized up to a critical temperature~$T_{\mathrm{co}}$.

In the absence of a magnetic field, these fluctuations have been studied in the superconducting phase
using the renormalization group (RG) method. \cite{emp} Integrating out step by step
the fast fluctuations around the mean-field solution~$u=\mathrm{i}\tau_2$ for superconductivity,
we reproduce the free energy functional~$\mathcal{F}$
while the coupling constants flow according to the RG equations
\begin{align}
 \frac{\mathrm{d}t}{\mathrm{d}\xi } &=
        \frac{3}{16\pi }t^{2}\ , \quad
 \frac{\mathrm{d}\ln(\kappa^{2}/t)}{\mathrm{d}\xi }
    = - \frac{3}{8\pi } t\ ,  \label{k4}
\end{align}
where $\xi$ is the running logarithmic variable of the RG
(see the Supplemental Material to Ref.~\onlinecite{emp}).

Equations (\ref{k4}) lead to a solution
for~$t$ increasing with~$\xi$ and one for~$\kappa$ that decreases.
The RG flow stops at $\xi = \ln(l_\mathrm{min}^{-1}/\kappa)$, yielding
\begin{align}
t &= t_{b}
  \Big[
     1 + \frac{3t_{b}}{16\pi }
         \ln\big(l_{\mathrm{min}}\kappa \big)
  \Big]^{-1},  \label{k4a} \\
\kappa^{2} &=
    \kappa_{b}^{2}
    \Big[
       1 + \frac{3t_{b}}{16\pi }
        \ln\big(l_{\mathrm{min}}\ \kappa \big)
    \Big]\ .  \label{k4b}
\end{align}
The ultraviolet cutoff in the logarithms is $l_{\mathrm{min}}\propto \lambda ^{-2}$,
the minimal length in the theory. The parameters~$t_{b}$ and~$\kappa _{b}^{2}$ denote the bare coupling constants,
determined by the actual values of temperature and curvature of the system.
The coupling constant~$\kappa _{b}^{2}$ constitutes a gap in
the excitation spectrum.

At small $\kappa$, the effective temperature $t$ becomes large, which should be interpreted
as phase transition from superconductivity to the disordered
pseudogap. Since simultaneously the effective anisotropy~$\kappa$ vanishes,
charge and SC suborders are
equally distributed in the pseudogap without a long-range order being formed.
The coupling constant $t$ diverges at the temperature
\begin{align}
  T_{c}= \alpha T^*\ \frac{1}{\ln(l_{\mathrm{min}}^{-1}/\kappa_b)}
 \label{k4c}
\end{align}
where $\alpha \sim 1$, cf. the paragraph after Eq.~(\ref{3a02}).
The transition temperature~$T_{c}$ can thus be considerably lower than $T^*$
where the pseudogap turns into the normal state.

In the presence of an external magnetic field, the critical temperature~$T_{c}$, Eq.~(\ref{k4c}),
does not change as long as the system remains in the Meissner state ($B<B_{c1}$). For $B>B_{c1}$,
vortices penetrate the sample, the ground state is no longer
homogeneous, and RG studies are more difficult.

At the same time, the RG scheme presented above is applicable to
the charge order--pseudogap transition since the ground state for $B > B_0$ [see Eq.~(\ref{k3})] is also homogeneous.
However, as easily seen, cf. Eq.~(\ref{3b02}), the gap in the excitation spectrum is now given by $\ell_{B}^{-2}-\kappa^{2}$, where
\begin{align}
 \ell_{B}= \sqrt{\frac{\hbar}{eB \sin \delta}} \gg l_{\mathrm{min}}
\label{k4d}
\end{align}
is the magnetic length. For the charge order state, we study fluctuations around the
mean-field solution~$u=\openone$, yielding the same RG equations~(\ref{k4}) and the
same solutions~(\ref{k4a}) and~(\ref{k4b}), yet
the coupling constant~$\kappa^2$ has to be replaced by the gap~$\ell_{B}^{-2}-\kappa^{2}$
that we treat as an effective coupling constant. As a result, we come
to the charge order--pseudogap transition temperature
\begin{align}
T_{\mathrm{co}}(B)
&=
\alpha T^*\ \frac{1}
     {\ln\big[l_{\mathrm{min}}^{-1}\ \big/\
      \big(\ell _{B}^{-2}-\kappa ^{2}\big)^{1/2}\big]}\ .  \label{k5}
\end{align}
For not too large magnetic fields~$B<2B_0$, $T_{\mathrm{co}}$ is
smaller than the superconducting transition temperature $T_{c}$, Eq.~(\ref{k4c}), while at $B = 2B_0$, we
have $T_{\mathrm{co}}(2B_0)=T_c$.
At larger~$B$, $T_{\mathrm{co}}$ grows slowly and the charge order--pseudogap critical line in the $T$-$B$
plane appears practically vertical. Formula~(\ref{k5}) actually is applicable down to the
critical field~$B_0$, Eq.~(\ref{k3}). Rewriting Eq.~(\ref{k5}) as
\begin{align}
B_{\mathrm{co}} = B_0 \ \Big(
1 + \frac{1}{l_\mathrm{min}^2\kappa_b^2}\ \exp\Big\{-\frac{2\alpha T^*}{T}\Big\}
\Big)
\label{k6}
\end{align}
yields a formula for the critical curve~$B_{\mathrm{co}}(T)$ valid for all temperatures and
the horizontal line in the $T$-$B$ plane at low~$T$ is evident.
On this line, the physics is characterized by degeneracy between the two suborders just as in the pseudogap at larger~$T$.
\bigskip

Equation~(\ref{k6}) leads to the $T$-$B$ phase diagram in Fig.~\ref{fig1}.
The boundary separating the SC and charge order phases is determined by
the asymptotic limit~(\ref{k3}) and is flat up to temperatures~$\sim T_{c}$, Eq. (\ref{k4c}).
The pseudogap state, characterized by the absence of charge or SC long-range orders, is located to the right of the phase
boundaries of the ordered regions but charge and SC suborders are as well equally distributed on the critical
line separating them. The SC critical temperature~$T_{c}$ at zero field agrees with the transition temperature~$T_{\mathrm{co}}$ at~$B=2B_0$.
Finally, the transition between SC and pseudogap at a finite magnetic field~$B<B_0$ is
qualitatively indicated by the dashed line in Fig.~\ref{fig1}, which also determines~$B_{c2}$.

%%%%%%%%%%%%%%%%%%%%%%%%%%%%%%%%%%%%%%%
%
% experiments, estimates
%
%%%%%%%%%%%%%%%%%%%%%%%%%%%%%%%%%%%%%%%

\section{Comparison with experiments}
Our conclusions agree with results of the recent experimental work \cite{leboeuf} by
LeBoeuf \emph{et al.}. Measuring
sound velocities in underdoped YBa$_{2}$Cu$_{3}$O$_{y}$ in
various directions up to magnetic fields of $30\ \mathrm{T}$, the authors encountered a thermodynamic transition
between superconductivity and a different state characterized by a biaxial charge order.
Referring to earlier NMR measurements \cite{wu}, they attributed this
state to a ``static charge'' order.

Following our results, we identify the critical magnetic field~$B_{\mathrm{co}}$
of Ref.~\onlinecite{leboeuf} with the field~$B_{\mathrm{co}}$ of Eq.~(\ref{k6}). Specifically at temperatures below
$T_{\mathrm{co}}\approx 40\ \mathrm{K}$, the field $B_{\mathrm{co}}$ was reported to be 
almost flat at $B_{\mathrm{co}}\approx 18\ \mathrm{T}$. This corresponds to
our theoretical low-temperature behavior, cf. Fig.~\ref{fig1}.
The right boundary of the charge order grows fast experimentally, which also is reflected in
and explained by the phase diagram and Eq.~(\ref{k6}) of our analysis. We mention
that the transition to charge ordering at $B\approx 28.5\ \mathrm{T}$ and $T=(50\pm10)\ \mathrm{K}$
as detected by NMR~\cite{wu} agrees
with our theoretical phase diagram as well. Furthermore, the
interpretation of the data in terms of the biaxial charge order supports the
theoretical prediction \cite{emp} of a checkerboard modulation.

Previous studies on the competition between superconductivity and another
order were typically based on Ginzburg-Landau-type (GL) theories with
the competing state typically chosen to be a spin-density wave (SDW) (see,
e.g., Refs.~\onlinecite{demler,kivelson,moon}). In such approaches, however,
it is not easy to explain the flat lower boundary of the ``static charge'' order in Fig.~2 of
Ref.~\onlinecite{leboeuf} because in GL theories the upper critical field $B_{c2}$, which should agree
with this boundary, essentially depends on temperature. Moreover, GL theories lack the emergence
of the pseudogap state, and the prediction of SDW competing with SC does not fit the
experimental evidence \cite{ghiringhelli,achkar,wu,leboeuf}. Finally, we mention that
the competition of superconductivity and a charge order has also been studied in
phenomenological models assuming interactions mediated by nearly critical charge collective modes. \cite{chargeorder_QCP}

It is useful to extract numerical estimates for the coupling constants in
the model~(\ref{3a02}) using the experimental data \cite{leboeuf}.
By Eq.~(\ref{k3}), the experimental value~$B_{0}\approx 18\ \mathrm{T}$
yields under the assumption~$\sin \delta =0.5$ that $\kappa ^{-1}\approx
9\ \mathrm{nm}$. Fitting formula~(\ref{k6}) to the experimental data of Ref.~\onlinecite{leboeuf} while imposing
the constraint~$B_{\mathrm{co}}(T_c)=2B_0$,
we extract the minimal length~$l_{\mathrm{min}} \approx 1.2\ \mathrm{nm}$.
Finally, formula~(\ref{k4c}) with~$T_c\approx 60.7\ \mathrm{K}$ determines the energy scale~$\hbar vS\approx 0.12\ e\mathrm{V}$.
Using $v\sim 2\cdot 10^{7}\ \mathrm{cm}\ \mathrm{s}^{-1}$
for the Fermi velocity \cite{fermivelocity}, we find~$S\sim 1\ \mathrm{nm}^{-1}$, i.e. the hot spots
take several $10\ \%$ of the Fermi surface.

These rough estimates fit our picture. In particular, the large ratio of
the gapped Fermi surface supports the idea of hot spots coalescing and forming
gapped regions that are centered at the antinodes and thus establish $d$-wave symmetry. \cite{chatterjee}
At the same time,
the estimates do not allow precise calculations of phase diagram parameters
such as $T^*$ or $T_{c}$ since in reality, the ratios
between these scales are close to unity ($T^*$ is just a few~$T_{c}$). Nevertheless, the
theoretical phase diagram in Fig.~\ref{fig1} illustrates well the experimental
situation as reported in Ref.~\onlinecite{leboeuf}.

%%%%%%%%%%%%%%%%%%%%%%%%%%%%%%%%%%%%%%%
%
% conclusion
%
%%%%%%%%%%%%%%%%%%%%%%%%%%%%%%%%%%%%%%%

\section{Conclusion}
Studying in the presence of a magnetic field the microscopically derived
theory of the pseudogap in high-$T_{c}$ cuprates from Ref.~\onlinecite{emp}, which
is characterized by an $\mathrm{SU}(2)$ order parameter fluctuating between
superconducting and charge order, we suggest an explanation
for the unusual phase diagram including both orders and the pseudogap
state observed experimentally in Ref.~\onlinecite{leboeuf}.

\acknowledgments

The work of K.B.E. has been supported by the ``Chaire Internationale de
Recherche Blaise Pascal'' financed by the State of France and the R\'egion
\^Ile-de-France. H.M. and K.B.E. acknowledge a financial support by the
SFB/TR12 of the Deutsche Forschungsgemeinschaft.
%H.M., M.E., and K.B.E. acknowledge the hospitality of the IPhT at the CEA-Saclay, where this work has been done.

%%%%%%%%%%%%%%%%%%%%%%%%%%%%%%%%%%%%%%%
%
% bibliography
%
%%%%%%%%%%%%%%%%%%%%%%%%%%%%%%%%%%%%%%%

%\bibliography{my_bib}

\begin{thebibliography}{99}

\bibitem{leenagaosawen}
P.~A.~Lee, N. Nagaosa, and X.-G. Wen,
Rev. Mod. Phys. \textbf{78}, 17 (2006).

\bibitem{Norman:605918}
M.~R.~Norman and C. P\'epin,
Rep. Prog. Phys. \textbf{66},  1547 (2003).

\bibitem{tranquada} J. M. Tranquada, B. J. Sternlieb, J. D. Axe, Y. Nakamura, and S. Uchida,
Nature, \textbf{375}, 561 (1995).

\bibitem{wise} W. D. Wise, M. C. Boyer, K. Chatterjee, T. Kondo, T. Takeuchi, H. Ikuta, Y. Wang, and E. W. Hudson,
Nat. Phys. \textbf{4}, 696 (2008).

\bibitem{fujita} K. Fujita, A. R. Schmidt,  Eun-Ah Kim, M. J. Lawler, D. H. Lee, J. C. Davis, H. Eisaki, and Shin-ichi Uchida,
J.~Phys.~Soc.~Jpn. \textbf{81}, 011005 (2012).

\bibitem{ghiringhelli}
    G. Ghiringhelli,
    M. Le Tacon,
    M. Minola,
    S. Blanco-Canosa,
    C. Mazzoli,
    N. B. Brookes,
    G. M. De Luca,
    A. Frano,
    D. G. Hawthorn,
    F. He,
    T. Loew,
    M. Moretti Sala,
    D. C. Peets,
    M. Salluzzo,
    E. Schierle,
    R. Sutarto,
    G. A. Sawatzky,
    E. Weschke,
    B. Keimer, and
    L. Braicovich,
Science \textbf{337}, 821 (2012).

\bibitem{achkar} A. J. Achkar, R. Sutarto, X. Mao, F. He, A. Frano, S. Blanco-Canosa, M. Le Tacon, G. Ghiringhelli, L. Braicovich, M. Minola, M. Moretti Sala, C. Mazzoli, R. Liang, D. A. Bonn, W. N. Hardy, B. Keimer, G. A. Sawatzky, and D. G. Hawthorn,
Phys. Rev. Lett. \textbf{109}, 167001 (2012).

\bibitem{chang} J. Chang,  E. Blackburn,	
A. T. Holmes,	
N. B. Christensen,	
J. Larsen,	
J. Mesot,	
Ruixing Liang,	
D. A. Bonn,	
W. N. Hardy,	
A. Watenphul,	
M. v. Zimmermann,	
E. M. Forgan,	
and S. M. Hayden,
Nat. Phys. \textbf{8}, 871 (2012).

\bibitem{blackburn} E. Blackburn, J. Chang, M. Hücker, A. T. Holmes, N. B. Christensen, Ruixing Liang, D. A. Bonn, W. N. Hardy, U. Rütt, O. Gutowski, M. v. Zimmermann, E. M. Forgan, and S. M. Hayden,
Phys. Rev. Lett. \textbf{110}, 137004 (2013).

\bibitem{DoironLeyraud:2007bj}
N. Doiron-Leyraud, C. Proust, D. LeBoeuf, J. Levallois, J.-B. Bonnemaison, R. Liang, D. A. Bonn, W. N. Hardy, and L. Taillefer,
Nature \textbf{447}, 565 (2007).

\bibitem{laliberte:2011}
F. Lalibert\'e,	
J. Chang,	
N. Doiron-Leyraud,	
E. Hassinger,	
R. Daou,	
M. Rondeau,	
B. J. Ramshaw,	
R. Liang,	
D. A. Bonn,	
W. N. Hardy,	
S. Pyon,	
T. Takayama,	
H. Takagi,	
I. Sheikin,	
L. Malone,	
C. Proust,	
K. Behnia, and	
L. Taillefer,
Nat. Commun.~\textbf{2}, 432 (2011).

\bibitem{Chang:2010} J. Chang, R. Daou, C. Proust, D. LeBoeuf, N. Doiron-Leyraud, F. Lalibert\'e, B. Pingault, B. J. Ramshaw, R. Liang, D. A. Bonn, W. N. Hardy, H. Takagi, A. B. Antunes, I. Sheikin, K. Behnia, and L. Taillefer,
Phys. Rev. Lett.~\textbf{104}, 057005 (2010).

\bibitem{Millis:2007wx}
A.~J. Millis and M. R. Norman,
Phys. Rev. B \textbf{76}, 220503 (2007).

\bibitem{Sebastian:2012vd}
S. E. Sebastian, N. Harrison, R. Liang, D. A. Bonn, W. N. Hardy, C. H. Mielke, and G. G. Lonzarich,
Phys. Rev. Lett. \textbf{108}, 196403 (2012).

\bibitem{Sebastian:2012ip}
S. E. Sebastian, N. Harrison, and G. G. Lonzarich,
Rep. Prog. Phys. \textbf{75}, 102501 (2012).

\bibitem{wu}  T. Wu,	
H. Mayaffre,	
S. Kr\"amer,	
M. Horvati\'c,	
C. Berthier,	
W. N. Hardy,	
R. Liang,	
D. A. Bonn, and
M.-H. Julien,
Nature~\textbf{477}, 191 (2011).

\bibitem{wu2013}
T. Wu,	
H. Mayaffre,	
S. Kr\"amer,	
M. Horvati\'c,	
C. Berthier,	
P. L. Kuhns,	
A. P. Reyes,	
R. Liang,	
W. N. Hardy,	
D. A. Bonn,
and M.-H. Julien, Nat. Commun.~\textbf{4}, 3113 (2013).

\bibitem{leboeuf}  D. LeBoeuf,	
S. Kr\"amer,	
W. N. Hardy,	
R. Liang,	
D. A. Bonn,
and C. Proust,
Nat.~Phys.~\textbf{9}, 79 (2013).

\bibitem{emp} K. B. Efetov, H. Meier, and C. P\'epin,
Nat. Phys.~\textbf{9}, 442 (2013).

\bibitem{acs} A. Abanov, A.V. Chubukov, and J. Schmalian,
Adv. Phys. \textbf{52}, 119 (2003).

\bibitem{ms2} M. A. Metlitski and S. Sachdev,
Phys. Rev. B \textbf{82}, 075128 (2010).

\bibitem{footnote01} While a controlled derivation of the $\sigma$-model~(\ref{3a02})
as effective low-energy theory formally assumes a small angle~$\delta$, cf. Ref.~\onlinecite{emp},
we may expect the at least qualitatively correct physical picture from this model also for~$\delta\sim 1$.

\bibitem{taillefer2013}
G. Grissonnanche,
O. Cyr-Choini\`ere,
F. Lalibert\'e,
S. Ren\'e de Cotret,
A. Juneau-Fecteau,
S. Dufour-Beaus\'ejour,
M.-E. Delage,
D. LeBoeuf,
J. Chang,
B. J. Ramshaw,
D. A. Bonn,
W. N. Hardy,
R. Liang,
S. Adachi,
N. E. Hussey,
B. Vignolle,
C. Proust,
M. Sutherland,
S. Kr\"amer,
J.-H. Park,
D. Graf,
N. Doiron-Leyraud,
and L. Taillefer, arXiv:1303.3856.

\bibitem{demler} E. Demler, S. Sachdev, and Y. Zhang,
Phys. Rev. Lett. \textbf{87}, 067202 (2001).

\bibitem{kivelson} S. A. Kivelson, Dung-Hai Lee, E. Fradkin, and V. Oganesyan,
Phys. Rev. B \textbf{66}, 144516 (2002).

\bibitem{moon} E. G. Moon and S. Sachdev,
Phys. Rev. B \textbf{80}, 035117 (2009).

\bibitem{chargeorder_QCP} S. Andergassen, S. Caprara, C. Di Castro, and M. Grilli,
Phys.~Rev.~Lett.~\textbf{87}, 056401 (2001);
S. Caprara, C. Di Castro, M. Grilli, and D. Suppa, \emph{ibid.}~\textbf{95}, 117004 (2005).


\bibitem{fermivelocity} G. Margaritondo
in \emph{The Gap Symmetry and Fluctuations in High-$T_c$ Superconductors},
edited by
J. Bok \emph{et al.} (Plenum Press, New York, 1998).

\bibitem{chatterjee} U. Chatterjee,
 M. Shi,
 D. Ai,
 J. Zhao,
 A. Kanigel,
 S. Rosenkranz,
 H. Raffy,
 Z. Z. Li,
 K. Kadowaki,
 D. G. Hinks,
 Z. J. Xu,
 J. S. Wen,
 G. Gu,
 C. T. Lin,
 H. Claus,
 M. R. Norman,
 M. Randeria,
 and J. C. Campuzano,
Nat.~Phys.~\textbf{6},~99~(2010).


\end{thebibliography}

\end{document}